\documentclass[aps,prl,twocolumn,showpacs]{revtex4}
\usepackage[dvips]{graphicx}
\usepackage{amsmath}
\usepackage{amsfonts}
\usepackage{amssymb}
\usepackage{dsfont} 
\usepackage[latin1]{inputenc}
\usepackage[english,frenchb]{babel}
\usepackage{ulem}

\begin{document}

\title{Sub-Poissonian atom number fluctuations using light-assisted collisions}

\author{Y.R.P.~Sortais, A.~Fuhrmanek, R.~Bourgain and A.~Browaeys}

\affiliation{Laboratoire Charles Fabry, Institut d'Optique, CNRS, Univ Paris Sud,
2 Avenue Augustin Fresnel,
91127 PALAISEAU cedex, France}

\date{\today}

\begin{abstract}
We investigate experimentally the number statistics of a mesoscopic ensemble of cold atoms in a microscopic dipole trap loaded from a magneto-optical trap, and find that the atom number fluctuations are reduced with respect to a Poisson distribution due to light-assisted two-body collisions. For numbers of atoms $N\gtrsim 2$, we measure a reduction factor (Fano factor) of $0.72\pm0.07$, which differs from $1$ by more than $4$ standard deviations. We analyze this fact by a general stochastic model describing the competition between the loading of the trap from a reservoir of cold atoms and multi-atom losses, which leads to a master equation. Applied to our experimental regime, this model indicates an asymptotic value of $3/4$ for the Fano factor at large $N$ and in steady state. We thus show that we have reached the ultimate level of reduction in number fluctuations in our system.
\end{abstract}

\maketitle

There is a growing interest in the study of mesoscopic systems containing between $10 - 100$ particles. For example, mesoscopic ensembles of ultra-cold atoms could be a useful tool for quantum information processing~(e.g. \cite{Lukin2001,SaffmanRMP2010}) or for precision measurements beyond the standard quantum limit~\cite{Wineland1992,Giovanetti2004}. They are also a test bed  for the investigation of many-body correlated quantum systems~\cite{BlochRMP2008} and for the study of collective effects such as super-radiance (e.g.~\cite{Akkermans2010}). All these applications
require the precise knowledge of the distribution of the number of atoms as the properties of these finite size samples are governed by their statistical nature. In particular the knowledge of the variance of the number distribution is important. In this paper we show experimentally and theoretically that one of the conceptually simplest mesoscopic systems, namely a few cold atoms in a tight dipole trap being loaded from a cold atomic cloud~\cite{Whitlock2009,Birkl2010,Fuhrmanek2010}, already exhibits non trivial sub-Poissonian statistics.

The preparation of an atomic sample with non-Poissonian atom number distribution requires a non-linear mechanism usually provided by interactions between ultra-cold atoms. For example the dispersive s-wave interaction was used to reduce the relative atom number fluctuations between the sites of a two-well potential~\cite{Oberthaler2010,Pritchard2007} or of an optical lattice~\cite{Kasevich2001}. This led in particular to the study of the Mott transition~\cite{Greiner2002,Gerbier2006}. This interaction was also used to demonstrate reduced atom-number fluctuations in a single tight dipole trap~\cite{Raizen2005}. Recently, the production of a sample with definite atom numbers was demonstrated using the Pauli blockade~\cite{Jochim2011}.

Inelastic collisions between ultra-cold atoms can also provide the non-linearity required to modify the atom number statistics, as shown recently~\cite{Steinhauer2010,Spreeuw2010}. In those experiments, three-body inelastic collisions induce losses in an initially trapped sample of $50-300$ atoms at or close to quantum degeneracy, and the ever decreasing fraction of remaining atoms exhibits reduced number fluctuations with respect to a Poisson distribution. Here, we consider theoretically a different and yet more general regime where the trap continuously experiences the interplay between a loading process from a reservoir of laser-cooled atoms and strong inelastic $\rho$-body losses ($\rho\geqslant 1$), and we investigate experimentally the case $\rho=2$  where the losses are due to light-assisted collisions. This situation is used elsewhere to, e.g., produce a single atom source~\cite{Schlosser2001,Weinfurter2006}. There, one operates in the ``collisional blockade" regime where the loading rate is such that the microscopic trap contains one or zero atom with equal probabilities ($\langle N\rangle=0.5$) and the atom number distribution is maximally sub-Poissonian with variance $\Delta N^2=0.5~\langle N\rangle$ ~\cite{Schlosser2002}.

Here, we explore the regime where $\langle N\rangle$ goes beyond $0.5$ in steady state as we increase the loading rate. In practice, we prepare a thermal ensemble of up to $10$ cold atoms at a temperature of $\sim100~\mu$K in a microscopic dipole trap. We observe that the atom number distribution remains sub-Poissonian and that the reduction in number fluctuations with respect to the Poisson distribution, $\Delta N^2/\langle N\rangle$, is locked to a constant value of $0.75$ for $\langle N\rangle\gtrsim 2$, a fact that was overlooked so far. To explain this fact, we use a microscopic approach that takes into account the stochastic nature of the competing loading and loss processes, and we calculate the atom number distribution at any time of the system evolution. We do so by solving a master equation both numerically and analytically, and find good agreement with the average result of a Monte Carlo approach where we study the individual behavior of atoms. Using this general theoretical approach, we analyze our data and find that we have reached experimentally the ultimate level of reduction in atom number fluctuations that one can expect in a dipole trap operating in our regime ($\rho=2$). The formalism presented in this paper is applicable to any system where a random loading process competes with a $\rho$-body loss process, whatever its nature and whatever the number of atoms.

To study the number statistics of a few atoms in the presence of competing random processes, we implemented the following experiment (details can be found in Ref.~\cite{Fuhrmanek2010}). First, we produced a microscopic optical dipole trap at $850$~nm by sharply focusing a laser beam~\cite{Sortais2007}. We then loaded this trap from a magneto-optical trap (MOT) of $^{87}$Rb atoms surrounding the region of the dipole trap. Atoms enter the dipole trap randomly, are trapped thanks to the cooling effect of the MOT beams, and are expelled from the trap due mainly to inelastic two-body collisions assisted by the near-resonant light of the cooling beams and, in a minor extent, to collisions with the residual background gas in the chamber (one-body losses). We measured elsewhere the two-body and one-body loss constants to be $\beta'\sim500$~(at.s)$^{-1}$ and $\gamma\sim0.2$~s$^{-1}$ respectively~\cite{Fuhrmanek2011}. The actual value of the loading rate $R$ is proportional to the MOT local density in the vicinity of the microscopic trap, which is the parameter that we vary. For values of $R\gg\beta'$ the mean number of trapped atoms in steady state exceeds unity and is $\langle N\rangle_{\textrm{st}} = \sqrt{R/\beta'}$, while for $R\ll\gamma$ it goes to zero as $\langle N\rangle_{\textrm{st}}=R/\gamma$. The intermediate regime corresponds to the ``collisional blockade" regime where $\langle N\rangle_{\textrm{st}}=0.5$. Experimentally we operate at $\langle N\rangle_{\textrm{st}}\gtrsim 2$ in the following.

\begin{figure}
\includegraphics[width=8cm]{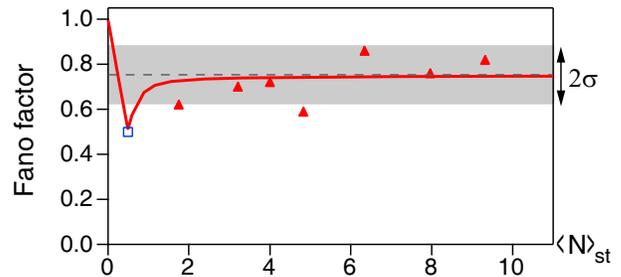}
\caption{Fano factor $F$ versus the average number of atoms $\langle N\rangle_{\textrm{st}}$ in the microscopic dipole trap in steady state. Square and triangles : experimental data collected in the ``collisional blockade" regime ($\langle N\rangle_{\textrm{st}}=0.5$) and beyond ($\langle N\rangle_{\textrm{st}} >0.5$). Solid line: model based on a stochastic process (see text) with $\beta'=500$~(at.s)$^{-1}$ and $\gamma=0.2$~s$^{-1}$. Dashed line : theoretical limit $F=3/4$ for $\langle N\rangle_{\textrm{st}}\gg 1$. $\sigma$: rms dispersion of the data collected beyond the collisional blockade regime.}
\label{Fig:FanoDataMicro}
\end{figure}

To get information on the number distribution of atoms in the dipole trap in steady state, we release the atoms from the trap and probe them with a pulse of resonant light. Using an intensifier to amplify single photon events above the noise of our CCD camera, we count the detected fluorescence photons individually~\cite{Fuhrmanek2010}. This number is proportional, on average, to the number of atoms $N$ in the trap before release. We build up the number distribution of counted photons by repeating this loading and probing experiment about $100$ to $1000$ times. Knowing the response of our imaging system to one atom exactly, we extract from the photon distribution the mean $\langle N \rangle_{\textrm{st}}$ and the variance $\Delta N^2$ of the atom number distribution in steady state, and calculate the corresponding Fano factor $F=\Delta N^2/\langle N \rangle_{\textrm{st}}$. The data shown in Fig.~\ref{Fig:FanoDataMicro} indicate a clear reduction of the atom number fluctuations with respect to a Poisson distribution for $\langle N \rangle_{\textrm{st}}\gtrsim 2$ with a mean $F=0.72$ and a total uncertainty ($1$ standard deviation) of $0.07$. This uncertainty is obtained by adding quadratically the statistical (type A) uncertainty of $0.05$ (deduced from the rms dispersion of the data) and the systematic (type B) uncertainty of $0.04$, which we estimated in previous work~\cite{Fuhrmanek2010}.

Qualitatively, this reduction can be understood as follows. If the losses were governed by random one-body events, e.g. background gas collisions, the trade-off between the random loading of the trap and the losses would result into a Poisson distribution with mean atom number $\langle N \rangle _{\textrm{st}}$ in steady state. If the losses now involve higher-body processes ($\rho\geq 2$) the loss rate varies as the number of $\rho$-uplets in the $N$-atom ensemble, i.e. increases non-linearly with $N$. For a given mean atom number this leads to the number distribution being narrower than a Poisson distribution, as the losses are more efficient on the high-$N$ side of the distribution.

To explain quantitatively the sub-Poissonian behavior of the atom number distribution, we use the following stochastic model that takes into account the three random processes involved, i.e the loading, the two-body losses, and the one-body losses. We consider the evolution in time of the probability $p_{N}(t)$ to have $N$ atoms in the dipole trap. To calculate the probability $p_N(t+dt)$, we sum the contributions of all channels associated to the random processes mentioned above that lead to having $N$ atoms in the trap at $t+dt$, given that the trap could possibly be filled with either $N-1$, $N$, $N+1$, or $N+2$ atoms at time $t$. The probability that a loading event occurs in the time interval $dt$ when there are already $N$ atoms in the trap is $R\,dt\,p_{N}(t)$. Similarly, the probability that a loss event occurs during $dt$ is $\gamma N\,dt\,p_N(t)$ for one-body events, and $\beta' \frac{N(N-1)}{2}\, dt\, p_{N}(t)$ for two-body events. We obtain eventually the following equation :
\begin{eqnarray}
p_{N}(t+dt) =& p_{N}(t)&(1-[R +\gamma N+\beta'\frac{N(N-1)}{2}]dt)\nonumber\\
& + p_{N-1}(t)& \,R\, dt\nonumber\\
& + p_{N+1}(t)& \, \gamma (N+1)\, dt\nonumber\\
& + p_{N+2}(t)& \, \beta' \frac{(N+2)(N+1)}{2}\, dt.\label{Eq:pn(t+dt)}
\end{eqnarray}
Taking the limit $dt\rightarrow 0$, Eq.~(\ref{Eq:pn(t+dt)}) yields the following master equation that rules the evolution of $p_N(t)$ in time:
\begin{eqnarray}
\frac{d p_N}{dt} = & R & (\mathds{E}^{-1}-\mathds{1})[p_{N}] \nonumber\\
& +~ \gamma &(\mathds{E}-\mathds{1})[Np_{N}] \nonumber\\
& +~ \beta' &(\mathds{E}^2-\mathds{1})[\frac{N(N-1)}{2}p_N],\label{Eq:MasterEquation}
\end{eqnarray}
where $\mathds{E}$ is the ``step operator" defined by its effect on an arbitrary function $f(N)$ :
\begin{equation}
\mathds{E}[f(N)] = f(N+1) ,~~~~\mathds{E}^{-1}[f(N)]=f(N-1),
\end{equation}
and $\mathds{1}$ is the identity operator. Using eqs.~(\ref{Eq:MasterEquation}) we obtain the equation of evolution of the mean number of atoms $\langle N\rangle=\sum\limits_{N=0}^\infty N\,p_N$:
\begin{equation}\label{Eq:diffeqmeannumber}
\frac{d\langle N\rangle}{dt}=R-\gamma\, \langle N\rangle-
\beta'\langle N\rangle(\langle N\rangle-1)-\beta'~\Delta N^2\ .
\end{equation}
When $\Delta N^2=0$, we recover the phenomenological equation sometimes used to describe the loading of a trap containing a small number of atoms~\cite{Ueberholz1999,Schlosser2002}, i.e. $dN/dt=R-\gamma~N-\beta'~N(N-1)$. When $\Delta N^2=\langle N\rangle$ (i.e. assuming a Poisson distribution), eq.~(\ref{Eq:diffeqmeannumber}) also yields the widely used equation $d\langle N\rangle/dt=R-\gamma~\langle N\rangle-\beta'~\langle N\rangle^2$. However, without any \textit{a priori} relation between $\Delta N^2$ and $\langle N\rangle$, eq.~(\ref{Eq:diffeqmeannumber}) cannot be solved analytically.

\begin{figure}
\includegraphics[width=8cm]{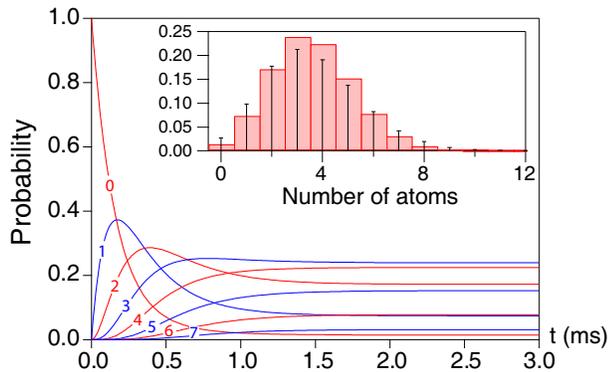}
\caption{Set of numerical solutions $\{p_{N}(t)\}$ of eqs.~(\ref{Eq:MasterEquation}), labeled by $N$. Parameters : $R=6000$\,s$^{-1}$, $\gamma=0.2$\,s$^{-1}$ and $\beta'=500$\,(at.s)$^{-1}$ lead to $\langle N\rangle_{\textrm{st}}=3.6\sim\sqrt{R/\beta'}$. Inset : filled bars : the set of solutions in steady state $\{p_{N}(t=3~\textrm{ms})\}$ (filled bars) is clearly sub-Poisson; sticks : Poisson distribution with same mean value.} \label{Fig:Evolution}
\end{figure}

\begin{figure}
\includegraphics[width=8cm]{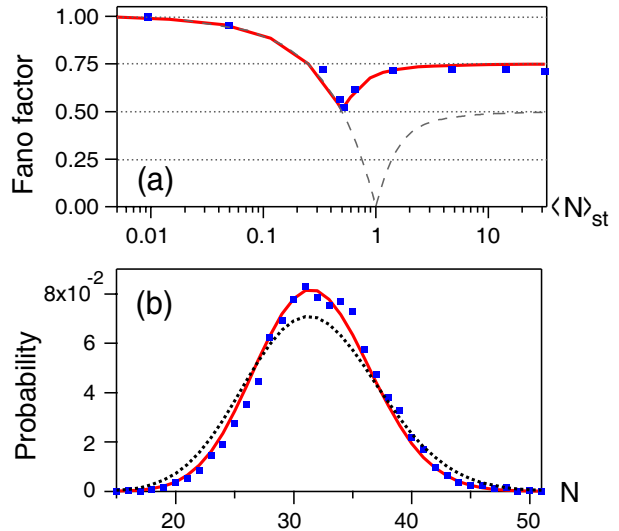}
\caption{Theoretical results obtained by different approaches. Solid line: numerical solution of eqs.~(\ref{Eq:MasterEquation}) for our experimental parameters ($\gamma=0.2$~s$^{-1}$, $\beta'\sim500$~(at.s)$^{-1}$); squares : Monte-Carlo simulation. (a) Dependence of the Fano factor on $\langle N\rangle_{\rm st}$. (b) Example of calculated atom number distribution in steady state, with $R=5\times10^5$~s$^{-1}$, yielding $\langle N\rangle_{\rm st}=32$. The numerical solution is indistinguishable from the Gaussian solution of eq.~(\ref{Eq:FokkerPlanck}). We find a Fano factor $F=3/4$. Dotted line : Poisson distribution with the same mean value. Dashed line in (a) : numerical solution of eqs.~(\ref{Eq:MasterEquation}) when two-body processes induce the loss of one atom only from the trap (parameters : $\gamma=5\times10^{-3}$~s$^{-1}$, $\beta'\sim500$~(at.s)$^{-1}$). In the limit $\gamma/\beta'\rightarrow 0$, $\langle N\rangle_{\rm st}$ can be locked to $1$ in a fully deterministic way ($\Delta N^2 = 0$).}\label{Fig:FanoTheory}
\end{figure}

To calculate the first moments of the number distribution, we used three different approaches. First, we solved numerically eqs.(\ref{Eq:MasterEquation}) using the boundary conditions $p_{N}(0)=\delta_{N,0}$ and, for $N\gg \langle N\rangle_{\rm st}$,  $p_{N}(t)=0$. As an example, Fig.~\ref{Fig:Evolution} illustrates the time evolution of the probabilities $p_N(t)$ for parameters leading to $\langle N\rangle_{\rm st}=3.6$. The number distribution is found to be sub-Poissonian, with $F=0.74$.
By varying the loading rate the same approach yields the distribution for any value of $\langle N\rangle_{\rm st}$. We analyze the case where $\gamma\ll \beta'$ in the following. When $\langle N\rangle_{\rm st}\ll 0.5$ we find, as expected, that the distribution is close to a Poisson law as one-body losses then dominate two-body losses (see Fig.~\ref{Fig:FanoTheory}(a))~\footnote{The distribution would remain Poissonian for any values of $\langle N\rangle_{\rm st}$ in the absence of two-body losses ($\beta'=0$), as can be derived analytically from eq.~(\ref{Eq:MasterEquation}).}. The presence of two-body processes induced losses of atom pairs leads to a sub-Poissonian behavior that is maximal for $\langle N\rangle_{\rm st}=0.5$, corresponding to $p_0=p_1=0.5$. While this regime has been described before (see Ref.~\cite{Schlosser2002}), the numerical approach predicts that atom number fluctuations do not become Poissonian for larger numbers of atoms. In fact, the Fano factor reaches an asymptotic value of $0.75$ as
soon as $\langle N\rangle_{\rm st}\gtrsim 2$, corresponding to a reduction of $-1.25$~dB with respect to the Poisson case. The numerical prediction  reproduces well our data, as shown in Fig.\,\ref{Fig:FanoDataMicro}.

The second approach to solve eqs.~(\ref{Eq:MasterEquation}) is analytical. It is valid for $\langle N\rangle_{\rm st}\gg 1$ only and follows closely the approach of Ref.~\cite{VanKampen}. We first re-write the master equation into a dimensionless rate equation :
\begin{eqnarray}
\frac{d p_N}{d\tau} = & \langle N\rangle_{\rm st} & (\mathds{E}^{-1}-\mathds{1})[p_{N}] \nonumber\\
& +~ \frac{1}{\langle N\rangle_{\rm st}} &(\mathds{E}^2-\mathds{1})[\frac{N(N-1)}{2}p_N],\label{Eq:MasterEquation_2}
\end{eqnarray}
where the one-body loss term of eq.~(\ref{Eq:MasterEquation}) has been neglected (following $R \gg\beta'\gg \gamma$) and $\tau=t\sqrt{R\beta'}$ is a dimensionless time variable. Since the number distribution in steady state is expected to be peaked around $\langle N\rangle_{\rm st}$ with a width on the order of $\sqrt{\langle N\rangle_{\rm st}}$, we consider the number of trapped atoms at time $\tau$ as a stochastic quantity of the form
\begin{equation}
N(\tau) = \langle N\rangle_{\rm st}~\phi(\tau)+ \sqrt{\langle N\rangle_{\rm st}}~\xi(\tau)
\end{equation}
where $\xi(\tau)$ is a stochastic variable with mean $\langle\xi\rangle(\tau)=0$ and an amplitude of $\sim1$, and $\phi(\tau)$, also on the order of $1$, is a deterministic and slowly varying function of time ($\phi(\tau)=\langle N\rangle(\tau)/\langle N\rangle_{\rm st}$). We then consider the probability $P(\xi,\tau)=p_N(\tau)$ that $N$ atoms are in the trap at time $\tau$. Since $p_{N+k}(\tau)=P(\xi+\frac{k}{\sqrt{\langle N\rangle_{\rm st}}},\tau)$ and $\langle N\rangle_{\rm st}\gg1$, we replace $p_{N+k}(\tau)$ in eq.(\ref{Eq:MasterEquation_2}) by a Taylor expansion of $P(\xi,\tau)$ in powers of $1/\sqrt{\langle N\rangle_{\rm st}}$. Replacing $p_N(\tau)$ by $P(\xi,\tau)$, the time derivative $dp_N/d\tau$ becomes $\partial_{\tau}P - \sqrt{N_{\rm st}}~\dot{\phi}~\partial_{\xi}P$~\footnote{Care must be taken when calculating the total derivative of $P(\xi,\tau)$ with respect to time, that eq.(\ref{Eq:MasterEquation_2}) was established with $N$ being held constant during an infinitesimal time interval $d\tau$, so that $\dot{\xi}(\tau)=-\sqrt{\langle N\rangle_{\rm st}}\dot{\phi}(\tau)$.}. Identification of the power terms in the expanded master equation then yields the following equations that rule the evolution of $\phi$ and $P$ in time :
\begin{eqnarray}
\dot{\phi}&=&1-\phi^2,\label{Eq:dphi/dt}\\
\partial_{\tau} P& =& 2\phi~\partial_{\xi}(\xi P)+ \frac{1}{2}(2\phi^2+1)~\partial^2_\xi P\label{Eq:FokkerPlanck} .
\end{eqnarray}
Eq.~(\ref{Eq:FokkerPlanck}) is a linear Fokker-Planck equation with time dependent coefficients, the steady state solution of which is Gaussian~\cite{VanKampen}. Finally, using eq.~(\ref{Eq:FokkerPlanck}) we find that $\langle\xi^2 \rangle$ evolves in time according to :
\begin{eqnarray}
\frac{d\langle \xi^2\rangle}{d\tau} = -4\phi \langle \xi^2\rangle +(1+2\phi^2).\label{Eq:evolution_de_<xi^2>}
\end{eqnarray}
Equations~(\ref{Eq:dphi/dt}) and (\ref{Eq:evolution_de_<xi^2>}) allow us to calculate the evolution of the Fano factor in time, $F(\tau)=\langle \xi^2\rangle(\tau)/\phi(\tau)$. In particular, in steady state, $\phi =1$ and $F=3/4$. This analytical finding is in excellent agreement with our numerical solution (see Fig.~\ref{Fig:FanoTheory}). Besides, we find that the analytical result is valid for atom numbers as small as $\sim 2$.

Finally, we cross-checked our theoretical results by a Monte-Carlo simulation, where we calculate at each time increment the survival probabilities of individual atoms to the various random events involved in the problem~\cite{Fuhrmanek2011}. By averaging over many atomic histories, we reconstructed atom number distributions and found Fano factors in very good agreement with those presented above (see Fig.~(\ref{Fig:FanoTheory})), which validates the master equation approach.

In conclusion, we discuss our experimental findings and theoretical approaches from a more general perspective. First, the observed reduction in number fluctuations is due to loss terms that vary non-linearly as $N^\rho$, and is thus intrinsically robust to losses (provided $\rho$ and the loading rate remain constant). In our case, we reached experimentally the ultimate level of reduction ($-1.25$~dB) predicted by theory when a loading mechanism competes with a two-body non-linearity leading to the loss of atom pairs, no matter the underlying mechanism (light-assisted collisions, hyperfine changing collisions...). More generally, the exact level of reduction achievable depends on $\rho$, on the number of atoms being lost after a $\rho$-body process, and on the presence (or the absence) of a loading mechanism. When $R\neq0$, the analytical approach explained above can be generalized and yields a Gaussian atom number distribution in steady state ($P(\xi,\tau)$ evolves according to a Fokker-Planck equation similar to eq.~(\ref{Eq:FokkerPlanck})). For $\rho$-body processes leading to losses of $\rho$-uplets, one finds an equation similar to eq.~(\ref{Eq:evolution_de_<xi^2>}) and $F=\frac{1}{2}(1+\frac{1}{\rho})$ in steady state. When $R=0$, slightly better levels of reduction can be achieved, as $F=\rho/(2\rho-1)$. This was recently demonstrated in the case of three-body losses~\cite{Spreeuw2010}. Finally, we extended our approach to the case where two-body collisions lead to the loss of one atom only from the trap~\footnote{We replaced $(\mathds{E}^2-\mathds{1})$ by $(\mathds{E}-\mathds{1})$ in eq.~(\ref{Eq:MasterEquation}).}, as is the case for elastic collisions induced evaporative losses and for some light-assisted loss mechanisms. Such mechanisms have been used recently to produce near-deterministically a single atom source for quantum information processing~\cite{Andersen2010}. Taking these mechanisms into account, our theoretical approach predicts that fluctuations fully vanish when only $1$ atom is left in the trap in the absence of one-body decay (see Fig.~\ref{Fig:FanoTheory}(a)), i.e. that a robust and fully deterministic preparation of single trapped atoms is in principle possible.

\begin{acknowledgments}
We acknowledge support from the E.U. through the ERC Starting Grant ARENA, and from IFRAF and Triangle de la Physique. A.~F. acknowledges partial support from the DAAD Doktorandenstipendium. We thank G.~Messin and M.P.A.~Jones for fruitful discussions.
\end{acknowledgments}

\end{document}